\documentclass[11pt]{article}

\usepackage[a4paper,margin=1.2in]{geometry}

\usepackage[utf8]{inputenc}
\usepackage{amsfonts,amsmath,amssymb,amsthm}
\usepackage{algorithm}
\usepackage{enumitem}
\usepackage{algpseudocode} 
\usepackage{verbatim}
\usepackage{url}
\usepackage{cite}

\newcommand{\Z}{{\mathbb Z}}

\newcommand{\Ort}{\mathcal{O}}
\newcommand{\A}{\mathcal{A}}

\newcommand{\Q}{\mathcal{Q}}

\newcommand{\ket}[1]{\lvert #1 \rangle}

\newcommand{\abs}[1]{\lvert #1 \rvert}

\newcommand{\braket}[2]{\langle #1 \vert #2 \rangle}

\DeclareMathOperator{\lcm}{\mathrm{lcm}}

\title{Secure multiparty quantum computations for greatest common divisor and private set intersection}
\author{
Muhammad Imran
\\
 Institute of Mathematics, Department of Algebra,
 \\
Budapest University of Technology and Economics,
\\
M\H{u}egyetem rkp. 3., Budapest, H-1111, Hungary.
\\
E-mail: \texttt{mimran@math.bme.hu
}}
\date{}
\date{}

\begin{document}

\maketitle

\begin{abstract}
We present a secure multiparty quantum computation (MPQC) for computing greatest common divisor (GCD) based on quantum multiparty private set union (PSU) by Liu, Yang, and Li. As the first step, we improve the security of the MPQC protocol for computing least common multiple (LCM) by Liu and Li by constructing an efficient exact quantum period-finding algorithm (EQPA) as a subroutine instead of the standard (probabilistic) Shor's quantum period-finding algorithm (QPA). The use of EQPA instead of the standard QPA guarantees the correctness of the protocol without repetitions. The improvement of LCM protocol also improves the private set union protocol which is based on computing LCM. Finally, using the same idea of the PSU protocol, we construct a quantum multiparty private set intersection (PSI) by transforming the PSI problem into the problem of computing GCD. Performance analysis shows that the correctness and the unconditional security in the semihonest model are guaranteed directly from the correctness and the security of the subroutine protocols (LCM and PSU protocols). Moreover, we show that the complexity of the proposed protocols is polynomial in the size of the secret inputs and the number of parties.

\end{abstract}

\section{Introduction}
Secure multiparty computation (MPC) is a subfield of cryptography with the goal of creating methods for parties to jointly compute a function over multiparty private inputs. Unlike traditional cryptographic tasks, where cryptography assures security and integrity of communication or storage where the adversary is outside the system, the cryptography in this model protects participants' privacy from each other. Since Yao’s seminal work \cite{yao1986generate} in 1982, MPC has attracted a lot of attention because it has a lot of important applications such as secret sharing, electronic voting, privacy-preserving computation, etc. On the other hand, as the field of quantum computing evolves, cryptography is one of the most influenced field. Quantum cryptography, which can be regarded as the quantum mechanics and classical cryptography, has been widely investigated on numerous branches such as quantum key distribution \cite{bennet1984quantum,ekert1991quantum,bennett1992quantum,cabello2000quantum,shih2009new}, quantum secret sharing \cite{hillery1999quantum,karlsson1999quantum,xiao2004efficient}, quantum key agreement \cite{zhou2004quantum,chong2011improvement,chong2010quantum}) including multiparty quantum computation (MPQC). It is an important and interesting question whether the principle of quantum computing can be applied into MPC. General purpose secure multiparty quantum computation was first studied by Cr{\'e}peau, Gottesman and Smith \cite{crepeau2002secure}. It is important to find efficient MPC protocols for some specific problems since this will also improve the performance of the general purpose MPC. 

%such as quantum key distribution \cite{bennet1984quantum,ekert1991quantum,bennett1992quantum,cabello2000quantum,shih2009new}, quantum secret sharing \cite{hillery1999quantum,karlsson1999quantum,xiao2004efficient}, quantum key agreement \cite{zhou2004quantum,chong2011improvement,chong2010quantum}. 

 The algorithms for computing greatest common divisor (GCD) and least common multiple (LCM) are important tasks in many cryptographic protocols, therefore it is very useful to have MPC protocols for the tasks. Moreover, Liu, Yang, and Li  in \cite{liu2023quantum} show that the availability of LCM protocol in quantum setting directly leads to quantum multiparty private set union by using Shor's factoring algorithm \cite{Shor}. The same idea can be used to construct quantum multiparty private set intersection having MPQC for GCD. Private set intersection (PSI) is a cryptographic primitive that allows two parties to learn the intersection of their input sets and nothing else. There has been a significant amount of work on privacy-preserving set operations, including set intersection as it has numerous applications which are not limited in cryptographic purposes such as testing human genomes \cite{shen2018efficient}, contact discovery \cite{demmler2018pir}, remote diagnostic \cite{brickell2007privacy}, record linkage \cite{he2017composing}, and many more. Most of the existing PSI protocols are based on traditional classical cryptosystems, which are proven to be vulnerable in quantum domain. This makes the requirement of quantum computer resistant PSI. Applying quantum cryptography in the design of PSI is an ideal approach to address these issues.  In order to construct quantum multiparty PSI in the quantum setting using similar approach for PSU in \cite{liu2023quantum}, an MPQC for GCD is required. However, it was still unclear how to construct an MPQC for GCD. According to the formula $\gcd(x,y)=\frac{xy}{\lcm(x,y)}$, one can obtain greatest common divisor by using both protocols for multiplication and LCM. However, the formula is only applicable to two integers and it is obvious that for the two-party case this is not secure since the two-party multiplication protocol always reveals each other inputs. Hence, the recursive generalization of the formula, i.e., $\gcd(a,b,c)=\gcd(a,\gcd(b,c))$, does not give any help to build secure protocol. A simple observation also shows that computing GCD cannot be done using the approach of \cite{li2022quantum} for LCM which is based on period-finding algorithm. Fortunately, the extension of LCM protocol to the private set union \cite{liu2023quantum} seems to be a promising method to construct a secure protocol for GCD. Specifically, we can transform the GCD problem to the private set union problem by working iteratively on the set of prime factors of the secret inputs.

\subsection{Our contributions}
In this paper, the first MPQC for computing greatest common divisor is proposed. The protocol is mainly based on the quantum multiparty PSU by Liu, Yang, and Li in \cite{liu2023quantum}. Since the PSU protocol is based on MPQC protocol by Li and Liu in \cite{li2022quantum}, firstly we revisited the protocol and improve the performance (success probability, security, and efficiency) of the protocol by constructing a new efficient exact quantum period-finding algorithm (EQPA) and use it as a subroutine instead of the standard (probabilistic) quantum period-finding algorithm (QPA). Finally, using the same idea of the PSU protocol, we construct a quantum multiparty private set intersection (PSI) by transforming the PSI problem into the problem of computing GCD. Concretely, we make the following contributions:
\begin{itemize}
    \item[1.] We present the first efficient exact quantum period-finding algorithm. The only sufficient information required is a multiple $m$ of the period $r$. Our assumption on having the information $m$ about the period $r$ is not standard. In fact, knowing a multiple of the period of the multiplicative group modulo the number $m$ would make it possible to factor $m$ in randomized classical polynomial time. However, there are some particular situations where a multiple of the period is known (or can be made available) such as finding period of elements in a finite field and constructing cryptographic protocols. The exact quantum period-finding algorithm modulo $m$ runs in time $\Ort(\log^4 m))$ while the standard Shor's period-finding algorithm has complexity $\Ort(\log\log m (\log^3m))$. The main idea of the algorithm is based on amplitude amplification method following the exact quantum algorithm for Simon's problem by Brassard and Hoyer \cite{BraHoy}.
    \item[2.] We propose a an improved version of the LCM protocol \cite{li2022quantum} using EQPA instead of the standard QPA. The proposed MPQC for LCM improves the security of Li-Liu's protocol. The total computation and communication complexity of the protocol are $\Ort(n^4m^4)$ and $\Ort(n^2m)$ respectively, where $n$ is the number of parties and $m$ is the size of the inputs while Li-Liu's computation complexity is $\Ort(n^3m^2)$ with the same communication complexity. However, considering the success probability of the standard QPA, Li-Liu's protocol needs $\Ort(\log (nm))$ repetitions. The repetition itself leads to some possible security issues. Therefore, the modified protocol is more secure when the number of participants and the size of the inputs grow as repetition itself can lead to some possible attacks. 

    \item[3.]Furthermore, we also propose an efficient secure MPQC for computing GCD and private set intersection. Specifically, all parties prepare the prime factors of their inputs by using Shor's factoring algorithm and then use the quantum multiparty private set union \cite{liu2023quantum} to jointly compute the union of all prime factors of their secret inputs. Finally, by using the voting procedure in \cite{li2022quantum} iteratively, they are able to obtain the greatest power of each prime factors that simultaneously dividing all the inputs and hence the GCD of their secret inputs is found. Finally, as a straight forward implication, we have a quantum multiparty PSI based on the GCD protocol using similar approach with the quantum multiparty PSU \cite{liu2023quantum}.
    \end{itemize}

\subsection{Outline}
The rest of the paper is organized as follows: In Section \ref{prem}, we briefly recall all the necessary tools and protocols for our results: Shor's factoring algorithm, Li-Liu's protocol for LCM, and the quantum multiparty private set union.
Section \ref{xshor} is fully devoted for exact quantum algorithms. In section \ref{amp}, we briefly discuss amplitude amplification which is a common technique used to derandomizing quantum algorithms. In Section \ref{eqpa}, we give a detailed construction of our exact quantum period-finding algorithm. Section \ref{MPQC} contains all the proposed MPQC protocols: an improved Li-Liu's protocol, the GCD protocol, and the private set intersection protocol. 

\section{Preliminary}\label{prem}
In this section, we give high level descriptions of Shor's factoring algorithm, Li-Liu's MPQC protocol for least common multiple, and the quantum multiparty private union by Liu, Yang, and Li. 
\subsection{Shor's factoring algorithm}
The well-known Shor’s factoring algorithm is able to factor any large integer $N$ efficiently. Shor’s factoring algorithm is based on a reduction of factoring to period-finding problem (observed by Miller in the 1970s). The main tool of Shor's factoring (to factor a large integer $N$) is the quantum period-finding algorithm (QPA) to find the period of the function $f:\Z\to \Z_N$ defined by $f(x)=a^x \bmod{N}$ (where $a$ is chosen at random), i.e., the smallest positive integer $r$ such that $f(x+r)=f(x)$. Quantum period-finding algorithm in modulo $N$ requires $\Ort((\log n)n^3)$ quantum operations, with $\Ort(\log n)$ uses of modular exponentiation where $n=\log N$ . The main subroutines of Shor’s period-finding algorithm are modular exponentiation and quantum Fourier transform. Modular exponentiation needs $\Ort(n)$ multiplications \cite{knuth1998art} and the Quantum Fourier Transform circuit is quadratic in $n$ \cite{Shor}. Hence, to find a factor of an odd number $N$, given quantum period-finding algorithm is as follows: choose a random $x\bmod{N}$ and find its period $r$ using the QPA. Finally, compute $\gcd(x^{r/2}-1,N)$. Since $(x^{r/2}-1)(x^{r/2}+1)=x^r-1=0\bmod{N}$, thus the $\gcd(x^{r/2}-1,N)$ fails to be a non trivial divisor of $N$ only for $r$ is odd. Hence, the procedure yields a non trivial divisor of $N$ with probability at least $1-1/2^{k-1}$, where $k$ is the number of distinct odd prime factors of $N$. The factoring process will be iterated over the obtained non trivial factors, then all prime factors of $N$ can be found.

\subsection{Li-Liu's MPQC for least common multiple}
\begin{paragraph}{Multiparty least common multiple problem:} Assume that there are $n$ parties: $P_1,\dots,P_n$, where each party $P_k$ has a secret integer $r_k\in \{0,1,\dots,2^m-1\}$. All $n$ parties want to jointly compute the $\lcm(r_1,\dots,r_n)$ without revealing their respective secret.
\end{paragraph}

The key idea of Li-Liu's protocol is based on the observation that given functions $f_1,\dots,f_n$ with period $r_1,\dots,r_n$ respectively, then the function $f(x)=(f_1(x),\dots,f_n(x))$ has period $r=\lcm(r_1,\dots,r_n)$. Thus, each party $P_i$ is equipped with the oracle of the secret function $f_i$ $(\ket{x}\ket{0}\mapsto\ket{x}\ket{f_i(x)}$)  and hence together they compute the superposition:
\[\frac{1}{\sqrt{N}}\sum_{x\in \Z_N}\ket{x}\ket{f_1(x)}\dots\ket{f_n(x)}\]
where $N=2^m$. Therefore, the period $r=\lcm(r_1,\dots,r_n)$ can be found by applying the quantum period-finding algorithm. However, because of the probabilistic nature of the QPA, an additional voting procedure is required to check the correctness of the QPA's output. Namely, each party votes whether the output divides their secret input. If the output divides all the secret inputs, then the output passes the verification. The voting procedure is based on the multiparty quantum summation by Shi \textit{et al.} in \cite{shi2016secure}.

The total computation and communication complexity of Li-Liu's protocol is $\Ort(n^3m^2)$ and $\Ort(n^2m)$ respectively. However, considering the success probability of the standard QPA, Li-Liu's protocol needs $\Ort(\log (nm))$ repetitions. A simple observation can show that the repetition itself can lead to some possible attacks specifically the parties can learn a factor of others in each repetition from the incorrect outputs and their own secrets. Hence, the risk increases as the repetition grows (the size $m$ of the inputs grows), especially in the malicious model.

\subsection{Quantum multiparty private set union}
\begin{paragraph}{Private set union problem:}
Assume that there are $n$ parties: $P_1,\dots,P_n$, where each party $P_i$ has a secret set $S_i\subseteq U$ where $U$ is the complete set of cardinality $N$: $2^{m-1}<N\leq 2^m$. All $n$ parties want to jointly compute the $\bigcup S_i$ without revealing their respective secret.
\end{paragraph}

The key idea of the quantum multiparty private set union proposed by Li, Yang, and Liu consists of three main steps: encoding procedure, an improved quantum multiparty computation for LCM, and decoding procedure. The encoding procedure transforms each of the secret set $S_i$ (for all $1\leq i\leq n$) to prime numbers and hence encode the set $S_i$ as the product of prime numbers image of all its elements. After the encoding procedure, the MPQC protocol for LCM (based on an improved QPA) is performed to find the LCM of all the encoded $S_i$. Finally, decoding procedure is done by (an improved) Shor's algorithm to get the union from the prime factors of the LCM obtained in the previous procedure.

The computation and communication complexity of the protocol are $\Ort(n^3m^3k^3\log(nmk))$ and $\Ort(n^2mk)$ respectively where $k$ is the upper bound of the cardinalities of the secret inputs $S_i$. The use of an improved QPA in the protocol increases the success probability of the LCM protocol to more than $99\%$ and hence eliminates the requirement of the necessary repetitions of Li-Liu's protocol. However, it is still interesting to have a deterministic protocol for the LCM to produce a correct output with certainty.

\section{Exact quantum algorithms}\label{xshor}
Shor's quantum algorithm \cite{Shor} can determine the order (period) of group elements efficiently, and it serves as the main tool for factoring integers. However, Shor's algorithm is polynomial-time in the expected sense, which means it may fail with a small probability and in the unlucky case may take a very long time to succeed, even may never terminate. The same case happens with Simon's algorithm \cite{Simon}. However, Brassard and Hoyer, in \cite{BraHoy}, came up with an exact quantum polynomial time for Simon's problem.  The Brassard-Hoyer algorithm utilizes a modified version of Grover's technique in \cite{Grover} to derandomize Simon's algorithm. Specifically, they propose a method that, assuming that we can construct a superposition in which the total squared amplitude of the "desired" constituents (intuitively, the probability of success) is $\frac{1}{2}$, boosts this success probability to $1$.

The question about the existence of exact quantum replacements for bounded quantum error probabilistic algorithms is a natural question, as it is analogous to derandomizing probabilistic classical algorithms. Besides, some earliest quantum algorithms that demonstrate the power of quantum computers, such as Deutsch-Jozsa procedure \cite{DeutschJozsa} and Bernstein-Vazirani problem \cite{BerVaz}, are exact. It is a difficult open question whether Shor's factoring algorithm can be derandomized. In \cite{MosZal}, Mosca and Zalka successfully derandomize Shor's algorithm for discrete logarithm problem in a cyclic group of known order. All previous exact quantum algorithms are uniform, which means the circuits for the algorithms can be classically computed in time polynomial in the logarithm of the inputs, see \cite{NishOzaUnif} for the details of uniform quantum circuits.

Here we consider the question whether Shor's period-finding algorithm can be derandomized in the uniform computational model assuming some knowledge. Note that we use the term order and period interchangeably (using the term order when we talk about group elements and the term period for general functions). As knowing a multiple of the order $\Z_m^*$ would factor $m$ in randomized classical polynomial time, finding orders of group elements with a known multiple of the order is not necessarily as hard as factoring, so a multiple of the period may be a good candidate for such a help. An important example where this help is available is the case of computing multiplicative orders (and testing primitivity) of elements of finite fields. Beside, this can be very useful for some cryptographic protocols, see section \ref{lcm}.

\subsection{Amplitude amplifications}\label{amp}
Amplitude amplification is a common technique used to boost up the success probability of quantum algorithms. The basic idea is to look at the final state of a quantum algorithm (before performing a measurement) as a state living in the plane generated by the good space (generated by all the desired outputs) and the bad space and then rotate it into the direction of the good space. Thus, amplitude amplification can be regarded as the generalization of Grover search algorithm \cite{Grover}. We present a brief review of the general amplitude amplification discussed by Brassad, Hoyer and Tapp in \cite{BraHoyTap}.

Given an algorithm $\A$ using no measurement, the amplitude amplification is a method to boost the success probability of the algorithm $\A$. On initial input $\ket{0}$, the algorithm $\A$ returns a pure superposition $\A \ket{0}=\sum_{i\in I}\ket{i}\ket{\Gamma_i}$ for some index set $I\subset \Z$. We consider $\chi:I \to \{0,1\}$ a Boolean function that separates the desired outcome states (all states $\ket{i}\ket{\Gamma_i}$ with $\chi(i)=1$) from the unwanted states (all states $\ket{i}\ket{\Gamma_i}$ with $\chi(i)=0$) as follows. Let $A=\{i\in I~|~\chi(i)=1\}$ and $B=\{i\in I~|~\chi(i)=0\}$. We write
$\A\ket{0}=\ket{\Gamma_a}+\ket{\Gamma_b}$, where
$$\ket{\Gamma_a}=\sum_{i\in A}\ket{i}\ket{\Gamma_i}\text{ and }\ket{\Gamma_b}=\sum_{i\in B}\ket{i}\ket{\Gamma_i}.$$
Hence the success probability of the algorithm $\A$ is $a=\braket{\Gamma_a}{\Gamma_a}=|\ket{\Gamma_a}|^2$. Therefore, the amplitude amplification operator for the algorithm $\A$ is defined as
\begin{equation}\label{eq1}
    \Q(\A, \chi, \phi, \varphi)=-\A S_0^{\phi}\A^{-1} S_{\chi}^{\varphi},
\end{equation}
where $S_{\chi}^{\varphi}$ and $S_{0}^{\phi}$ are phase changing operators defined by
$$S_{\chi}^{\varphi}\ket{i}\ket{\Gamma_i}=\left\{\begin{array}{ll}
\varphi\ket{i}\ket{\Gamma_i}
& \mbox{~if $\chi(i)=1$} \\
\ket{i}\ket{\Gamma_i} & \mbox{~otherwise,}
\end{array}\right.~~~~\text{and}~~~~S_{0}^{\phi}\ket{i}\ket{\Gamma_i}=\left\{\begin{array}{ll}
\phi\ket{i}\ket{\Gamma_i}
& \mbox{~iff $i=0$} \\
\ket{i}\ket{\Gamma_i} & \mbox{~otherwise,}
\end{array}\right.$$
with $\phi$ and $\varphi$ are complex number of unit length.

The operator $\Q$ is a generalization of Grover's iterations applied in his quantum search algorithm \cite{Grover}. Moreover, by setting $\phi=\varphi=-1$, we have for every $j\geq 0,$
$$\Q^j \A\ket{0}= k_j\ket{\Gamma_a}+l_j\ket{\Gamma_b}$$
where $$k_j=\frac{1}{\sqrt{a}}\sin ((2j+1)\theta) ~~~~~\text{and}~~~~ l_j=\frac{1}{\sqrt{1-a}}\cos((2j+1)\theta),$$
and $0 \leq \theta\leq \pi/2$ is defined so that $\sin^2\theta=a=|\ket{\Gamma_a}|^2$.

A natural question to ask whether it is possible to boost the success probability to certainty. It turns out there are positive answers to this question. In \cite{BraHoy}, Brassard and Hoyer present an optimal value for the parameters $\phi$ and $\varphi$, namely $\phi=\varphi=\sqrt{-1}$, such that whenever the success probability of an algorithm $\A$ is $\frac{1}{2}$, then one application of the amplitude amplification $\Q$ boosts the success probability to $1$. This is the approach that Brassard and Hoyer use to derandomize Simon's algorithm. Another positive answer is also presented in \cite{MosZal} by Mosca and Zalka. They use one application of $\Q$ with parameters $\phi=\varphi=-1$ to increase the success probability $\frac{1}{4}$ of an algorithm $\A$ to $1$. They use this variant of amplitude amplification to present an exact quantum Fourier transform and derandomize Shor's quantum algorithm for discrete logarithm over groups of known orders. Therefore, one application of the exact quantum Fourier proposed by Mosca and Zalka requires three applications of the usual quantum Fourier transform.

As one may notice from some previous derandomizations, such as Simon's algorithm and Shor's discrete logarithm, the knowledge of the success probability of the algorithms makes the derandomizations possible. Therefore, in section \ref{eqpa}, we show that a multiple of the unknown order is sufficient to adjust the success probability to $\frac{1}{2}$. Hence,  this amplitude amplification derandomizes Shor’s order finding algorithm when a multiple of the order is known.

\subsection{Exact quantum period-finding algorithm}\label{eqpa}

The problem we consider is given a function $f$ with a promise that there exists a period $r$ such that $f(x)=f(y)$ if and only if $x=y\bmod{r}$, and a multiple $N$ of the unknown period $r$, determine the period $r$.  The first part of the algorithm is the standard Fourier sampling. We use here an exact version based on the exact quantum Fourier transform of Mosca and Zalka \cite{MosZal}. The standard Fourier sampling procedure maps $\ket{0}\ket{0}$ to $\sum_{k=0}^{m-1}\ket{k}\ket{\Gamma_k}$, where $\ket{\Gamma_k}=\frac{1}{m}\sum_{j=0}^{m-1}\omega^{kj}\ket{f(j)}$ and $\omega=e^{2\pi i/m}$.
Write $j$ as $j_0+rj_1$ ($0\leq j_0\leq r-1$).
Then

$$\ket{\Gamma_k}=
\left\{\begin{array}{ll}
1/r\sum_{j_0=0}^{r-1}\omega^{kj_0}\ket{f(j_0)}
& \mbox{~if $m/r$ divides $k$;} \\
0 & \mbox{~otherwise,}
\end{array}\right.
$$ whence
$${\abs{\Gamma_k}}^2
=\left\{\begin{array}{ll}
1/r & \mbox{~if $m/r$ divides $k$;} \\
0 & \mbox{~otherwise.}
\end{array}\right.
$$
In words, we have terms with $\ket{k}$ in the first register
only for those $k$ which are multiples of $m/r$.
Initially, any $k$ which is nonzero modulo $m$
is useful because $\frac{m}{\gcd(k,m)}$ is a proper divisor
of $r$. 
We have $\sum_{k\neq 0}\abs{\Gamma_k}^2=1-\frac{1}{r}$. However, fortunately, if we already know a divisor $d$ of $r$ then those values $k$ that give us new information are the non-multiples
of $\frac{m}{d}$. We have
$\sum_{kd\neq 0}\abs{\Gamma_k}^2=1-\frac{d}{r}$. The point is we do not know $r$.

The second part of the algorithm is based on the discussion in the last part of the previous paragraph. We maintain a divisor $d$ of $r$. We construct iterations of a procedure that increase $d$. Initially $d:=1$. As long as $d<r$, we find $k$ such that $dk\bmod{m}\neq 0$. Then we replace $d$ with $\frac{m}{gcd(m,k)}$ since this is another divisor of $r$ greater than $d$. Hence, $d$ keeps increasing as long as $d<r$ and it stops immediately when $d=r$ as $dk=0\bmod{m}$ for all $k$ if and only if $d$ is a multiple of $r$.

In order to construct an exact algorithm for the iteration procedure above, we need to adjust the probability to $\frac{1}{2}$ of each iteration as follows. Assume $d<r$. Let $rep(dk)$ be the smallest positive integer representative of $dk\bmod{m}$. In this case, $rep(dk)=d\frac{m}{r}$ for all $k$. Then $rep(dk)$ divides $m$ and all the $\frac{m}{rep(dk)}-1=\frac{r}{d}-1$ positive integers of the form $t  d\frac{m}{r}<m$ are the nonzero multiple of $d\frac{m}{r}$ modulo $m$. Note that if $\frac{r}{d}$ is even, then the integers of the form $t d\frac{m}{r}$ with $m/2\leq t  d\frac{m}{r}<m$ represent just half of multiples of $d\frac{m}{r}$ modulo $m$. However, if $\frac{r}{d}$ is odd, we need to add another multiple of $d\frac{m}{r}$ modulo $m$, say $d\frac{m}{r}$, with weight $\frac{1}{2}$. The problem is we do not know $d\frac{m}{r}$. However, fortunately, for at least one integer $0\leq j \leq \log_2m$, namely for $j=\lceil \log_2d\frac{m}{r}\rceil$, the interval $(0, 2^j]$ contains only $d\frac{m}{r}$ and no other multiple of $d\frac{m}{r}$ as if $j-1<\log_2d\frac{m}{r}\leq j$ then $d\frac{m}{r}\leq 2^j$ and $2d\frac{m}{r}>2^j$. 

Based on the descriptions above, we summarize the exact algorithm in the following pseudocode. 

\begin{algorithm}[H]\label{alg1}
\caption{Exact quantum period-finding algorithm}
\scriptsize
\begin{algorithmic}[1]
\State \textbf{Initialize:} $d \gets 1$, ${Found}\gets 1$;
\While{${Found > 0}$}
    \For{$j=-1,\ldots, \lfloor \log_2 m\rfloor$}\vspace{0.1cm}
	   \State{$\chi_j(k,b)=
		\left\{\begin{array}{ll}
			1 & \mbox{if $rep(dk)\geq\frac{m}{2}$ or
					$b=1$ and $0<rep(dk)\leq 2^j$;} \\
			0 & \mbox{otherwise;}
		\end{array}
		\right.$}
	\State{${\mathcal U}_j:
  		\ket{0}\ket{0}\ket{0}\ket{0}\mapsto 
\ket{\psi_j}=\frac{1}{\sqrt{2}}\sum
\ket{k}\ket{\Gamma_k}\ket{b}\ket{\chi_j(k,b)}$;\mbox{~~~~~~~~}}\Comment{where $k\in \{0,1,\dots, m-1\}$, $b\in \{0,1\}$.\mbox{~~~~~~}} 
	\State{Apply the amplitude amplified version of ${\cal U}_j$ to obtain$\ket{\psi'_j}=\sum
c'_{\chi_j(k,b)}\ket{k}\ket{\Gamma_k}\ket{b}\ket{\chi_j(k,b)}$.}
	\State{Look at the $\ket{k}$-register;}
	\If{$dk\neq 0\bmod{m}$} 
		\State{$d\gets \frac{m}{gcd(m,k)}$;}
    \Else
        \State{${Found}\gets {Found}-1$}
        \EndIf
    \EndFor
\EndWhile
\end{algorithmic}
\end{algorithm}
Each round consists of iterations for $j=-1,\dots,\lfloor \log_2m\rfloor$ instead of starting with index $j=0$ to cover both cases when $\frac{r}{d}$ is even and when $\frac{r}{d}$ is odd. The case when $\frac{r}{d}$ is even is covered at least once, when $j=-1$ where the interval $(0,2^j]$ does not contain any integer. While the case when $\frac{r}{d}$ is odd is covered at leat once, when $j=\lceil \log_2d\frac{m}{r}\rceil$.

As in each round before termination, the size of $d$ is increased by at least a factor $2$ and it stops immediately when $d=r$, we need at most $\lceil \log_2 r \rceil$ rounds of iterations. The overall number of calls to the exact Fourier transform or its inverse is $\Ort(\log m\log r)=\Ort(\log^2m)$.

\section{Proposed MPQC protocols}\label{MPQC}

\subsection{The MPQC least common multiple based on EQPA}\label{lcm}
The goal is to replace the standard QPA by our EQPA in Liu-Yang-Li's protocol. Therefore, we want to make sure that the requirement of EQPA is fulfilled, namely we provide a multiple of the least common multiple or the period of the common function $f(j)=f_0(j)||\dots||f_{n-1}(j)$. In step $(1)$, each party $P_i$ chooses a random $q$ such that $x_iq\sim 2^m$ and sends $y_i=x_iq$ to $P_0$. Therefore, $P_0$ has a multiple of the least common multiple by computing $k=\prod_{i=0}^{n-1} y_i$ and broadcasts it to all parties. Moreover, each party $P_i$ is equipped with the period function $f_i:\Z_k\to \Z_k$ defined by $f_i(x)=x\bmod{r_i}$. The rest of the protocol follows the original Liu-Yang-Li's protocol but using EQPA instead the standard QPA. We give the summary of the protocol in algorithm 3.
\noindent\rule{15cm}{0.7pt}
\begin{itemize}
    \item[\textbf{(1)}] For each $P_i$, chooses a random $q\in [2^l]$ such that $x_iq \cong 2^m$ and sends $y_i=x_iq$ to $P_0$.
    \item[\textbf{(2)}]  $P_0$ computes computes $k=\prod_{i=0}^{n-1} y_i$ and broadcasts it to all parties.
    \item[\textbf{(3)}]  For $0\leq i \leq n-1:$ each $P_i$ holds the function $f_i:\Z_k\to \Z_k$ be $f_i(x)=x\bmod{r_i}$.
    \item[\textbf{(4)}]  For $P_0$:
    \begin{itemize}
        \item[(a)] prepares two $m$-qubit quantum registers $h,t$ initialized as $\ket{0}_h\ket{0}_t$;
        \item[(b)] applies $H^{\otimes m}$ on $h$:
        \[
        \ket{0}_h\ket{0}_t \mapsto \frac{1}{\sqrt{k}}\sum_{j\in [k]}\ket{j}_h\ket{0}_t;
        \]
        \item[(c)] applies $CNOT^{\otimes m}$ on $h,t$, where $h$ controls $t$:
        \[
       \frac{1}{\sqrt{k}}\sum_{j\in [k]}\ket{j}_h\ket{0}_t \mapsto \frac{1}{\sqrt{k}}\sum_{j\in [k]}\sum_{j\in [k]}\ket{j}_h\ket{j}_t;
        \]
        \item[(d)] prepares an $k$-qubit quantum register $e_0$ initialized as $\ket{0}_{e_0}$;
        \item[(e)] applies $U_{f_0}: \ket{j}_t\ket{0}_{e_0}\mapsto \ket{j}_t\ket{f_0(j)}_{e_0}$ on $t, e_0:$
        \[
        \frac{1}{\sqrt{k}}\sum_{j\in [k]}\ket{j}_h\ket{j}_t\ket{0}_{e_0} \mapsto  \frac{1}{\sqrt{k}}\sum_{j\in [k]}\ket{j}_h\ket{j}_t\ket{f_0(j)}_{e_0};
        \]
        \item[(f)] sends $t$ to $P_1$.
    \end{itemize}
    \item[\textbf{(5)}]  For $P_i$, $1\leq i \leq n-1$:
    \begin{itemize}
        \item[(a)] prepares an $m$-qubit registers $e_i$ initialized as $\ket{0}_{e_i}$;
        \item[(b)] applies $U_{f_i}:\ket{j}_t\ket{0}_{e_i}\mapsto \ket{j}_t\ket{f_i(j)}_{e_i}$ on $t, e_i:$
        \[
        \frac{1}{\sqrt{k}}\sum_{j\in [k]}\ket{j}_h\ket{j}_t\ket{f_0(j)}_{e_0}\ket{f_(j)}_{e_1}\dots \ket{f_{i-1}(j)}_{e_{i-1}}\ket{0}_{e_i} \]
        \[ \mapsto
       \frac{1}{\sqrt{k}}\sum_{j\in [k]}\ket{j}_h\ket{j}_t\ket{f_0(j)}_{e_0}\ket{f(j)}_{e_1}\dots \ket{f_{i-1}(j)}_{e_{i-1}}\ket{f_i(j)}_{e_i};
        \]
        \item[(c)] sends $t$ to $P_{i+1}$.
    \end{itemize}
    \item[\textbf{(6)}]  For $P_0$:
    \begin{itemize}
        \item[(1)] applies $CNOT^{\otimes m}$ on $h,t$, where $h$ controls $t$:
        \[ 
        \frac{1}{\sqrt{k}}\sum_{j\in [k]}\ket{j}_h\ket{j}_t\ket{f(j)}_e \mapsto  \frac{1}{\sqrt{k}}\sum_{j\in [k]}\ket{j}_h\ket{0}_t\ket{f(j)}_e, 
        \]
        where $f(j)=f_0(j)||\dots||f_{n-1}(j)$, $e=(e_0,\dots, e_{n-1})$;
        \item[(2)] measures $t$, if $t$ is not $\ket{0}$, then rejects, otherwise continues;
        \item[(3)] Applies EQPA to find the period $r$ of $f$;
        \item[(4)] Broadcasts $r$ to all other parties.
    \end{itemize}
\end{itemize}
\noindent\rule{15cm}{0.7pt}

\begin{paragraph}{Correctness proof.}
The correctness of the protocol is ensured by the property of EQPA being deterministic and the fact that the function $f(x)=(f_0(x),\dots,f_{n-1}(x))$ has period $r=\lcm_{i=0}^n(r_i)$. 
\end{paragraph}

\begin{paragraph}{Security analysis.}
    In the first step, each $P_i$ sends $y_i=x_iq$ to $P_0$. However, $P_0$ cannot gain any useful information as $y_i$ is a multiplication of the secret input $x_i$ with a random element $q$. Moreover, following the security analysis of Li-Liu's protocol \cite{li2022quantum}, the protocol is secure under the three possible attacks (\textit{direct measurement attack, pre-period-finding attack, post-period-finding attack}) in the semihonest model. In the malicious model, our protocol seems more secured compared to Li-Liu's protocol because there is no repetition of the protocol is required.
\end{paragraph}

\begin{paragraph}{Complexity analysis.}
Note that the parties share a multiple $k$ of each of their secret so, $k$ is a multiple of the LCM. Moreover, since $k=\Ort(2^{mn})$, the most time consuming step of the protocol is the EQPA procedure that has $\Ort(\log^4 k)=\Ort(m^4n^4)$ computational complexity. Therefore, the total computational communication complexity are $\Ort(m^4n^4)$ and $\Ort(m^2n)$
\end{paragraph}

\subsection{The proposed MPQC for GCD and private set intersection}
\subsubsection{Multiparty quantum computation for GCD.}
  Assume that there are $n$ parties: $P_0,\dots,P_{n-1}$, where each party $P_k$ has a secret integer $r_k\in \{0,1,\dots,2^m-1\}$. All $n$ parties want to jointly compute the $\gcd(r_1,\dots,r_n)$ without revealing their respective secret. Furthermore, assume that the communication process is done via an authenticated quantum channel.

\begin{itemize}
    \item[\textbf{(1)}] For $P_i$, $0\leq i \leq n-1$ : applies Shor's factoring algorithm to obtain the set $R_i$ of all prime factors of $r_i$
    
    \item[\textbf{(2)}] All parties jointly perform the private set union protocol to get the set $R=\bigcup_{i=0}^{n-1}R_i$. 
    \item[\textbf{(3)}] For each prime $p\in R$, do the following iteration: using the multiplication protocol, all parties jointly vote whether $p, p^2,\dots$ divide their secret inputs in order to get the largest power $p^k$ that simultaneously divides all their secret inputs. Thus, the GCD can be obtained by the product of all the largest prime power of all elements of $R$.

\end{itemize}
\noindent\rule{15cm}{0.7pt}
\begin{paragraph}{Correctness proof.}
In the first step, each party performs Shor's factoring on their inputs to get the set of all prime factors of $r_i$. Therefore, each party can easily verify that they hold a correct set of prime factors of their inputs before applying private set union protocol in the next step. Since the correctness of the second step follows directly from \cite{liu2023quantum}, then it is left to show that the last step indeed gives the gcd of the secret inputs $r_i$'s. The last step indeed gives a correct output according to the definition of greatest common divisor
\[\gcd(p_1^{a_1}\cdots p_m^{a_m}, p_1^{b_1}\cdots p_m^{b_m})=p_1^{\max\{a_1,b_1\}}\cdots p_m^{\max\{a_m,b_m\}}\]
which is true for computing GCD for any $n$ numbers through their prime factorizations. Note that the success probability of \cite{liu2023quantum} is greater than $99\%$ but still probabilistic. Using the EQPA in the subroutine can guarantee the output with certainty.
\end{paragraph}

\begin{paragraph}{Security analysis.} Since there is no meaningful information can be gained regarding the secret inputs from the set $R$, then the security of the protocol follows directly from the security of the private set union protocol \cite{liu2023quantum}. Since the private set union protocol is unconditionally secure in the semihonest model, then similar security holds for the proposed multiparty quantum computation for GCD.
    
\end{paragraph}
\begin{paragraph}{Complexity analysis.}
The use of Shor's factoring in the first step of the protocol costs $\Ort(nm^2\log m)$ computational complexity. On the other hand, the computational and communication complexity of the private set union are $\Ort(n^3m^3k^3\log(nmk))$ and $\Ort(n^2mk)$ respectively where $k$ is the upper bound of the cardinality of the sets $R_i$'s. Thus, the second step has $\Ort(n^3m^6\log(nm^2))$ computational complexity and communication complexity $\Ort(n^2m^2)$. As for the last step, there are at most $m$ iterations of voting procedure, thus the computational and communication complexity of the last step are $\Ort(nm^3)$ and $\Ort(nm^2)$ respectively. Hence the total computational and communication complexity are $\Ort(n^3m^6\log(nm^2))$ and $\Ort(n^2m^2)$ respectively. On the other hand, using the EQPA to get a deterministic output in the subroutine of the PSU protocol gives extra computational complexity with total computational complexity $\Ort(n^4m^6\log(nm^2))$ instead of $\Ort(n^3m^6\log(nm^2))$.
\end{paragraph}

\subsubsection{Multiparty quantum private set intersection}
Assume that there are $n$ parties: $P_1,\dots,P_n$, where each party $P_i$ has a secret set $S_i\subseteq U$ where $U$ is the complete set of cardinality $N$: $2^{m-1}<N\leq 2^m$. All $n$ parties want to jointly compute the $\bigcup S_i$ without revealing their respective secret. Furthermore, assume that the communication process is done via an authenticated quantum channel. The protocol for private set intersection straightforwardly follows the protocol for private set union by Liu, Yang, and Li. We  
give the key steps of the protocol as follows:
\begin{itemize}
    \item[\textbf{(1)}] \textbf{Encoding: } each party $P_i$ transforms the elements of their corresponding secret set $S_i$ into prime numbers and encodes the secret $S_i$ as the product of all primes representation of its elements. 
    
    \item[\textbf{(2)}] \textbf{GCD protocol: } apply the multiparty quantum computation for GCD to compute the greatest common divisor of all the encoded $S_i$. 
    \item[\textbf{(3)}] \textbf{Decoding: } use the improved Shor's factoring algorithm in \cite{liu2023quantum} to factor the GCD obtained in the previous step and get the intersection of all $S_i$ from the prime factors of the GCD.
\end{itemize}
\noindent\rule{15cm}{0.7pt}
\begin{paragraph}{Correctness proof.}
The correctness of the protocol follows directly from the correctness of the GCD protocol and the fact that the prime factors of the greatest common divisor are common prime factors of all the encoded $S_i$. Hence decoding the prime factors give the elements of the intersection of all sets $S_i$'s. The same case with the GCD protocol, the original version of PSI protocol gives more than $99\%$ of success probability and we can get the certainty by using the EQPA in the subgroutines. 
\end{paragraph}

\begin{paragraph}{Security analysis.}
The security of the protocol follows as well directly from the security of the GCD protocol. Thus, it follows the security of the private set union \cite{liu2023quantum} which has unconditional security in the semihonest model.
\end{paragraph}

\begin{paragraph}{Complexity analysis.}
The most computational costs comes from the GCD protocol which is the same complexity with the PSU protocol which is $\Ort(n^3m^6\log(nm^2))$. In order to get the correct output with certainty, we can use the EQPA instead which increases the total complexity becomes $\Ort(n^4m^6\log(nm^2))$. While the communication complexity remains $\Ort(n^2m^2)$.
\end{paragraph}

\bibliographystyle{unsrt}
\bibliography{exshor}
\end{document}